\documentclass[%
 reprint,
 superscriptaddress,
 amsmath,
 amssymb,
 amsfonts,
 aps,
prb,
]{revtex4-2}

\usepackage[english]{babel}

\usepackage{graphicx}
\usepackage{hyperref}

\usepackage{mathptmx}

\begin{document}

\title{Constraints on a split superconducting transition under uniaxial strain in \texorpdfstring{Sr\textsubscript{2}}\texorpdfstring{RuO\textsubscript{4}} from scanning SQUID microscopy }

\author{Eli Mueller}
\affiliation{Stanford Institute for Materials and Energy Sciences, SLAC National Accelerator Laboratory, 2575 Sand Hill Road, Menlo Park, California 94025, USA}
\affiliation{Department of Physics, Stanford University, Stanford, California 94305, USA}

\author{Yusuke Iguchi}

\affiliation{Stanford Institute for Materials and Energy Sciences, SLAC National Accelerator Laboratory, 2575 Sand Hill Road, Menlo Park, California 94025, USA}
\affiliation{Geballe Laboratory for Advanced Materials, Stanford University, Stanford, California 94305, USA}

\author{Christopher Watson}
\affiliation{Stanford Institute for Materials and Energy Sciences, SLAC National Accelerator Laboratory, 2575 Sand Hill Road, Menlo Park, California 94025, USA}

\author{Clifford W. Hicks}
\affiliation{Max Planck Institute for the Chemical Physics of Solids, N{\"o}thnitzer Stra{\ss}e 40, Dresden 01187, Germany}
\affiliation{School of Physics and Astronomy, University of Birmingham, Birmingham B15 2TT, UK}

\author{Yoshiteru Maeno}
\affiliation{Department of Physics, Graduate School of Science, Kyoto University, Kyoto 606-8502, Japan}
\affiliation{Toyota Riken - Kyoto University Research Center (TRiKUC), Kyoto University, Kyoto 606-8501, Japan}

\author{Kathryn A. Moler}
\affiliation{Stanford Institute for Materials and Energy Sciences, SLAC National Accelerator Laboratory, 2575 Sand Hill Road, Menlo Park, California 94025, USA}
\affiliation{Department of Physics, Stanford University, Stanford, California 94305, USA}
\affiliation{Geballe Laboratory for Advanced Materials, Stanford University, Stanford, California 94305, USA}

\begin{abstract}
More than two decades after the discovery of superconductivity in Sr$_2$RuO$_4$, it is still unclear whether the order parameter has a single component or two degenerate components. For any two-component scenario, application of uniaxial strain is expected to lift the degeneracy, generating two distinct phase transitions. The presence of a second (lower-temperature) transition may be observable by probes that are sensitive to changes in the London penetration depth, $\lambda$, as a function of temperature, $T$. Here, we use scanning SQUID microscopy combined with a uniaxial strain device to test for a second transition under strain. We only observe a single transition. Within the temperature range where a second transition has been suggested by $\mu$SR measurements \cite{grinenko2021split}, we further place a tight upper bound of less than 1\% on the change in the zero temperature superfluid density $n_s\propto\lambda^{-2}(0)$ due to a second transition, suggesting that such a transition does not occur. These results constrain theories of the order parameter in Sr$_2$RuO$_4$.
\end{abstract}

\maketitle

\section{Introduction}
\label{sec:Intro}

Strontium ruthenate (Sr$_2$RuO$_4$) has attracted considerable scientific interest as an unconventional superconductor \cite{maeno2001intriguing,mackenzie1998extremely,mackenzie2020personal,mackenzie2003superconductivity,liu2015unconventional}. After more than a quarter century since its discovery \cite{maeno1994superconductivity}, the symmetry of the superconducting order parameter in Sr$_2$RuO$_4$ continues to be debated \cite{leggett2021symmetry,mackenzie2017even}. Measurements probing the spin part of the order parameter have ruled out spin-triplet pairing \cite{chronister2021evidence, pustogow2019constraints,ishida2020reduction}. What remains unclear is the symmetry of the orbital part of the order parameter. Evidence for time-reversal symmetry breaking (TRSB) superconductivity comes from muon spin relaxation ($\mu$SR) measurements\cite{luke1998time,grinenko2021split,grinenko2021unsplit,grinenko2023mu}, non-zero Kerr rotation below the superconducting critical temperature $T_c$ \cite{xia2006high}, and signatures in junction experiments that suggest chiral domain formation in the superconducting state \cite{anwar2013anomalous,nakamura2012essential}. In addition, ultrasound measurements have provided evidence of a two-component order parameter, consistent with chiral superconductivity \cite{ghosh2021thermodynamic,benhabib2021ultrasound}. Table \ref{tab:table1} summarizes the even-parity order parameters allowed under the point group $D_{4h}$ for the tetragonal symmetry of Sr$_2$RuO$_4$. The only order parameter that can account for TRSB superconductivity with a symmetry protected, two-component  degeneracy is chiral $d_{xz} \pm id_{yz}$ pairing \cite{suh2020stabilizing,clepkens2021higher}. However, given the two-dimensional nature of the Fermi surface, an order parameter with horizontal line nodes would be unexpected. In addition, thermal conductivity \cite{hassinger2017vertical} and scanning tunneling microscopy measurements \cite{sharma2020momentum} indicate vertical line nodes. Therefore, alternative two-component order parameters, $d \pm ig$ and $d \pm is$, have been proposed which assume that the system is tuned so that two one-dimensional pairing states are accidentally degenerate \cite{kivelson2020proposal,clepkens2021higher,wang2022higher,sheng2022multipole,romer2021superconducting,clepkens2021shadowed,romer2019knight}.

\begin{figure}
 \centerline{\includegraphics[width=\linewidth]{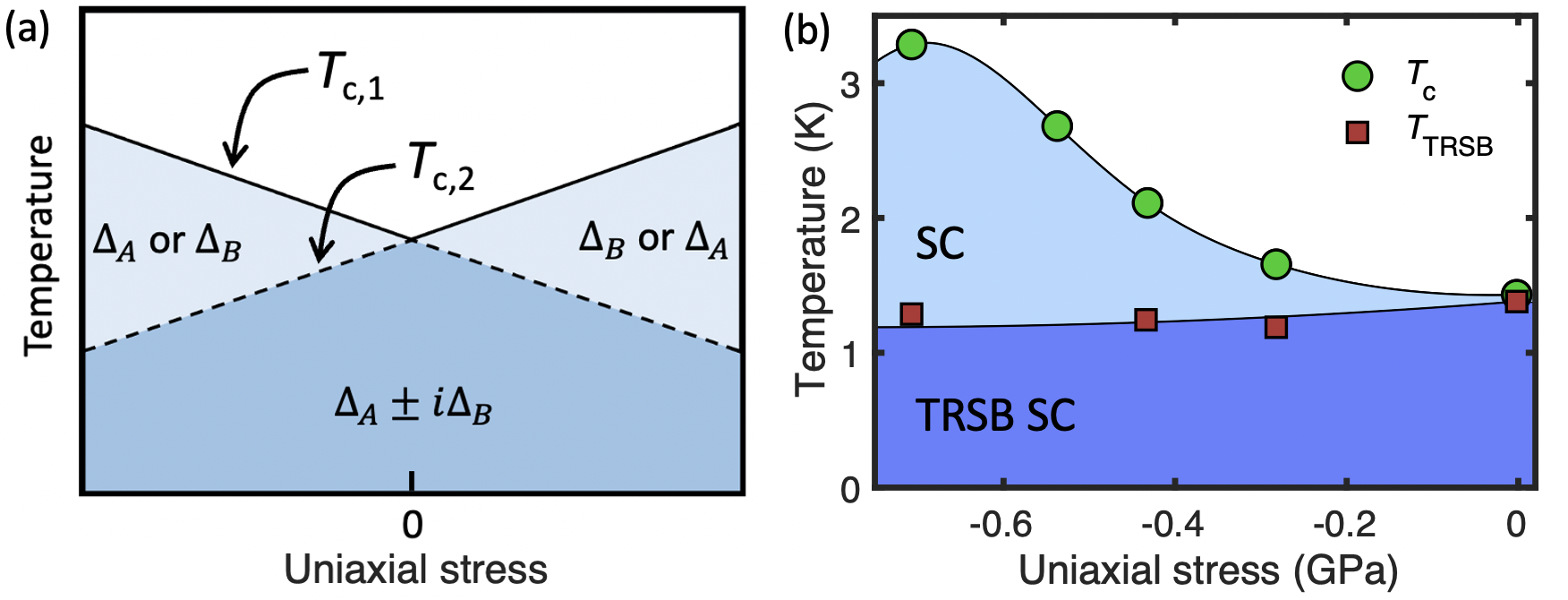}}
  \caption{Phase diagram for a two-component TRSB superconductor under uniaxial stress. (a) Schematic of the stress-temperature phase diagram for a general two-component ($\Delta = \Delta_A \pm i \Delta_B$) TRSB superconductor with a transition splitting under uniaxial stress. (b) Experimentally observed $T_c$ and $T_{\text{TRSB}}$ versus uniaxial compression along the $\langle100\rangle$ crystal axis (data adapted from \cite{grinenko2021split}). The onset of TRSB below $T_c$ has been interpreted as evidence of a $\{d_{xz},d_{yz}\}$ to $d_{xz} + id_{yz}$ transition of the order parameter.}
  \label{fig:hypothesis}
\end{figure}

For a general two-component, TRSB order parameter, $\Delta = \Delta_{A} + i\Delta_{B}$, application of a uniaxial stress is expected to lift the degeneracy of the $\Delta_{A}$ and $\Delta_{B}$ components and create two distinct phase transitions at $T_{c,1}$ and $T_{c,2}$ with a cusp in the $T_{c,1}$ versus strain curve as shown in Fig. \ref{fig:hypothesis}(a). For the case of chiral $d_{xz} + id_{yz}$ pairing, the degeneracy is protected by the tetragonal symmetry of the lattice and the splitting arises as the applied strain breaks this symmetry. However, for the $d + ig$ and $d+ is$ states, the degeneracy is not symmetry protected and requires fine tuning to achieve $T_{c,1} \approx T_{c,2}$ in the unstressed system. Applying an external strain alters this fine tuning and produces a transition splitting because the strain affects the $T_c$'s of the two components differently.

\begin{table}[b]
\caption{\label{tab:table1}
Even-parity (spin-singlet) irreducable representations of the tetragonal point group $D_{4h}$ \cite{sigrist1991phenomenological}.}
\begin{ruledtabular}
    \begin{tabular}{l c c r}
    \textrm{Irrep.} & \textrm{Order Parameter} & \textrm{Dimensionality} & \textrm{Nodal Structure}\\
    \colrule
        $A_{1g}$ & $s$ &  1 & None\\
        $A_{2g}$ & $g_{xy(x^2-y^2)}$  & 1 & Vertical Line \\
        $B_{1g}$ & $d_{x^2-y^2}$ &  1 & Vertical Line \\
        $B_{2g}$ & $d_{xy}$ &  1 & Vertical Line \\
        $E_{g}$ & $\{d_{xz},d_{yz}\}$ &  2 & Vert. \& Horiz. Line \\
    \end{tabular}
\end{ruledtabular}
\end{table}

Experimental evidence for a transition splitting in uniaxially strained Sr$_2$RuO$_4$ is ambiguous. Bulk and local susceptibility measurements under strain show a smooth quadratic dependence of $T_c$ near zero strain \cite{hicks2014strong,steppke2017strong,watson2018micron}. Furthermore, heat capacity and elastocaloric effect measurements show no evidence of a second transition under strain \cite{li2021high,li2022elastocaloric}. However, $\mu$SR measurements show an apparent splitting between $T_c$ and the onset of TRSB, $T_{\text{TRSB}}$ [Fig. \ref{fig:hypothesis}(b)] \cite{grinenko2021split}. The discrepancy between the non-observations and the $\mu$SR results have called into question whether the TRSB observed in $\mu$SR and Kerr measurements is associated with a bulk transition of the superconducting state.


If the anomaly at $T_{\text{TRSB}}$ corresponds to a transition in the superconducting order parameter, then this transition should be accompanied by an anomaly in the superfluid density and may be observable by probes that are sensitive to the London penetration depth. Bulk mutual inductance measurements on the multi-component superconductor UPt$_3$ have indeed shown two transitions that coincide with the double jump seen in heat capacity \cite{signore1995inductive,schottl2000magnetic,bruls1990strain}, however the magnitude of $\Delta\lambda$ associated with the second transition was not reported in these studies. Multiple superconducting phases are theoretically expected in the hexagonal lattice symmetry of UPt$_3$ at ambient pressure. In contrast, a very homogeneously strained state would be required to probe any clear evidence of a possible second transition in uniaxially strained Sr$_2$RuO$_4$. Bulk measurements on Sr$_2$RuO$_4$ under strain \cite{hicks2014strong,steppke2017strong,li2021high} show significant broadening of the superconducting transition due to inhomogeneity of the applied strain, which may obscure any signature of a second transition below $T_c$ \cite{roising2022heat}.

To test for the presence of a second transition, $T_{c,2}$, we perform scanning superconducting quantum interference device (SQUID) microscopy on a single crystal of Sr$_2$RuO$_4$ under uniaxial compression along the $\langle100\rangle$ crystal axis. In our microscope, the local diamagnetic response of a superconductor is probed with micron-scale spatial resolution. By measuring the local magnetic susceptibility versus temperature under fixed compression, we place an upper limit of less than 1\% on the contribution to the zero temperature superfluid density $\lambda^{-2}(0)$ from a second order parameter with a transition in the temperature range suggested by $\mu$SR data \cite{grinenko2021split}.

\section{Methods}
\label{sec:methods}

\begin{figure}
 \centerline{\includegraphics[width=\linewidth]{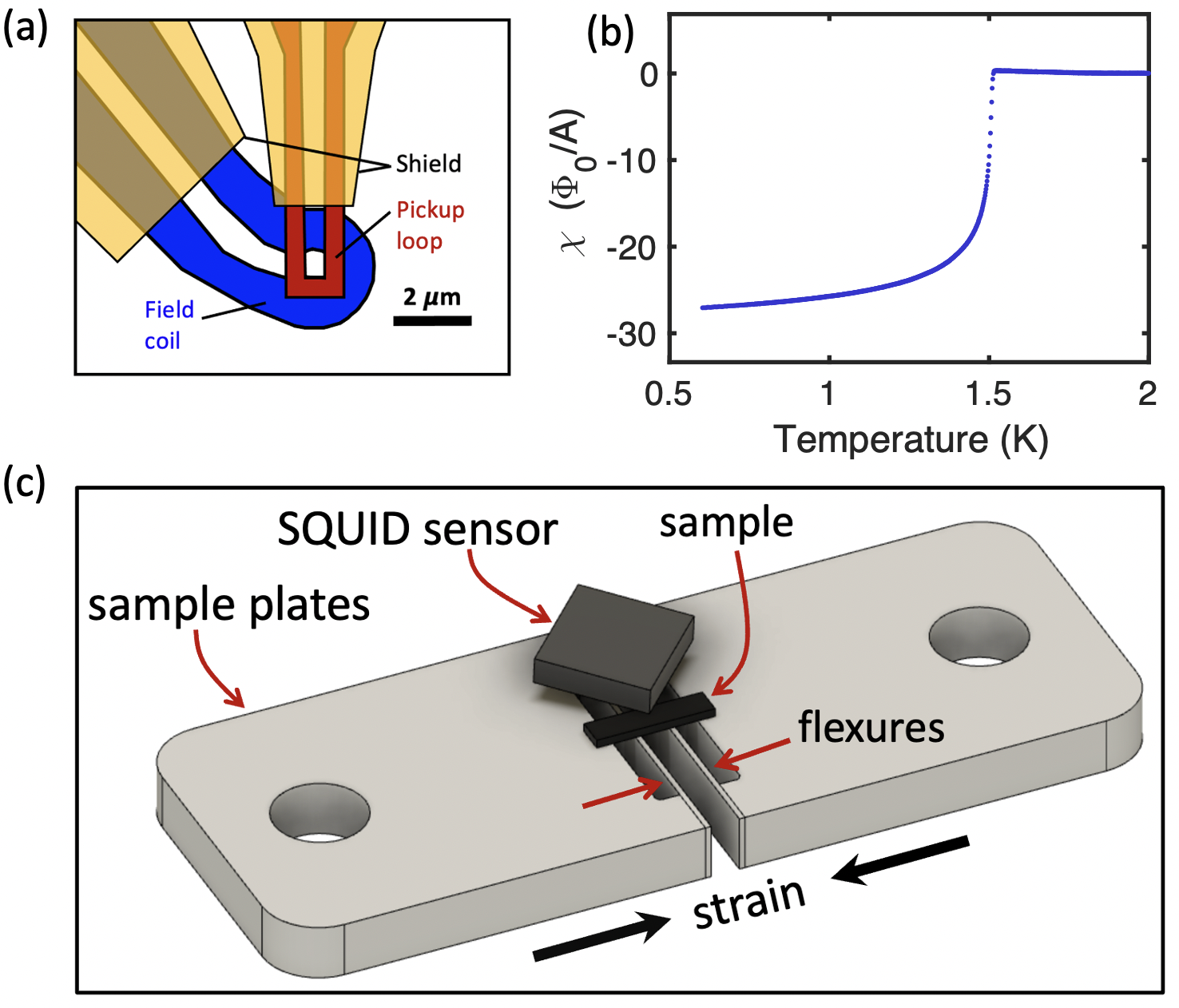}}
  \caption{Experimental setup for scanning SQUID microscopy under uniaxial strain. (a) Schematic of the SQUID susceptometer pickup loop (red) and field coil (blue). Superconducting shields (yellow) minimize magnetic flux coupling into the SQUID circuit. (b) Local susceptibility as a function of temperature measured under zero applied strain, showing a sharp onset of diamagnetism at the critical temperature. (c) Schematic of the movable sample plates with a sample mounted for uniaxial stress and a SQUID sensor chip aligned to the sample. The sample length is approximately 2~mm. The holes shown are used to anchor the sample plates to the main body (not shown) of the strain device. Deformation of the main body via piezoelectric actuators induces a displacement of the sample plates. The flexures are incorporated to reduce sample bending under strain and are made of titanium plates with a thickness of 100~$\mu$m.}
  \label{fig:methods}
\end{figure}

The main components of our SQUID susceptometers \cite{kirtley2016scanning} consist of a magnetic flux sensing pickup loop with 0.25 $\mu$m inner radius and a concentric field coil with 0.5 $\mu$m inner radius as shown in Fig. \ref{fig:methods}(a). We apply an ac current with amplitude $I_{\text{fc}}$ through the field coil, generating an ac magnetic flux through the pickup loop. Near a superconducting sample, the Meissner effect screens the field generated by the field coil and reduces the mutual inductance of the pickup loop-field coil pair. By recording this mutual inductance, we obtain a measure of the local magnetic susceptibility of the sample, $\chi$, which we report in units of $\Phi_0/A$, where $\Phi_0 = h/2e$ is the flux quantum. To measure the temperature dependence of the local susceptibility at a point on the sample, we bring the SQUID sensor chip in light mechanical contact with the sample and cycle the temperature while recording the susceptibility. An example of such a measurement at zero applied strain is shown in Fig. \ref{fig:methods}(b) where a sharp onset of diamagnetism ($\chi < 0$) is observed near the $T_c \approx$ 1.5~K. 

\begin{figure*}
 \centerline{\includegraphics[width=\linewidth]{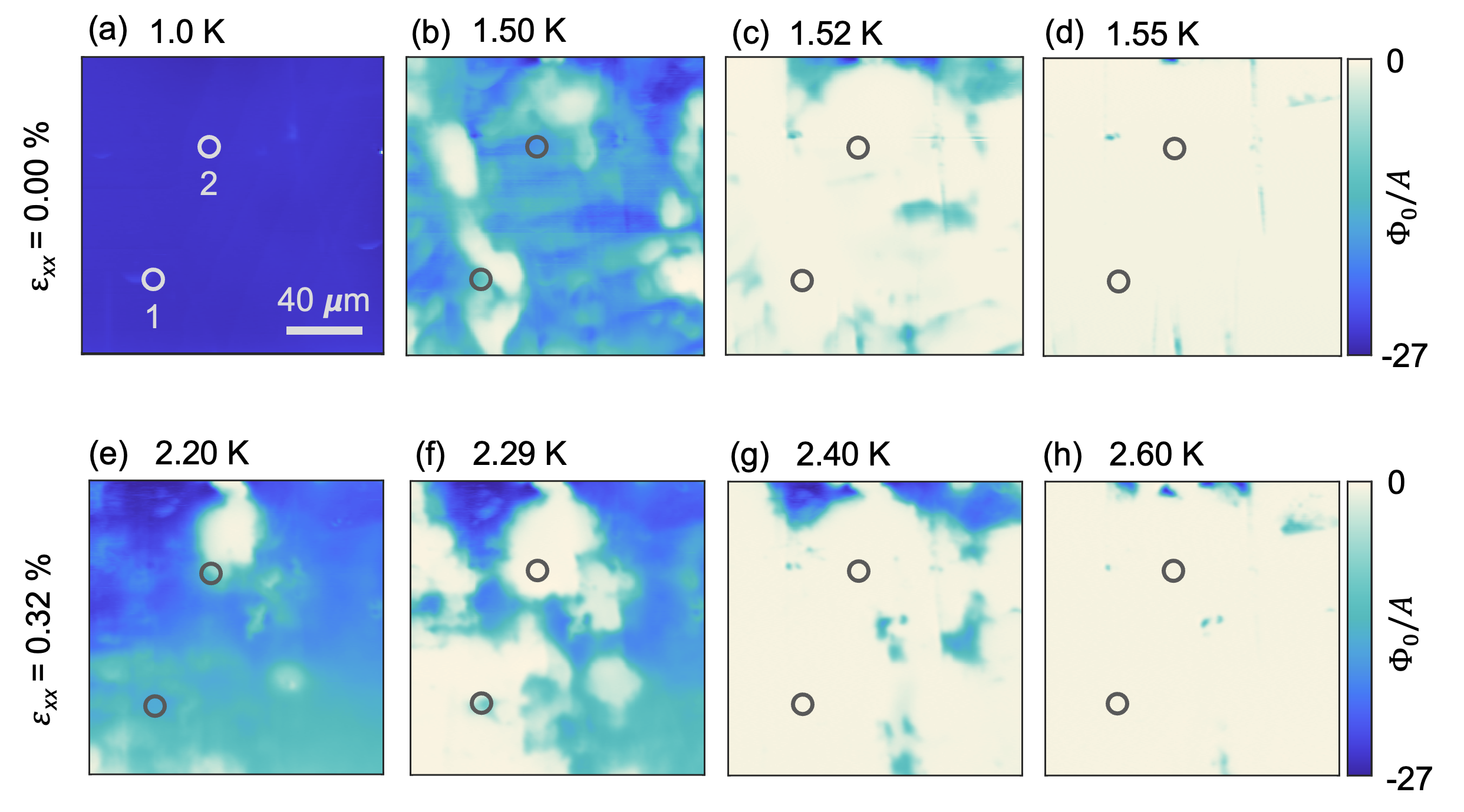}}
  \caption{Spatial inhomogeneity of diamagnetism through the bulk $T_c$ (a)-(d) Temperature series scans of susceptibility under zero applied strain. At $T$ = 1.0 K, the sample is superconducting indicated by a nearly uniform diamagentic susceptibility. The markers indicate the points that local susceptibility versus temperature was measured under a series of compressive strains. At temperatures near the bulk $T_c=$ 1.5~K, spatial variations in susceptibility indicate small inhomogeneity in local $T_c$. At $T$ = 1.55~K, the majority of the sample is in the normal state. Several small spots of weak diamagnetism are likely due to impurities and defects causing strain enhanced $T_c$ locally. (e)-(h) Susceptibility scans over the same region with an applied strain of $\epsilon_{xx} = -$0.32\%. Under this compression the bulk $T_c$ over this region increases to $\sim$2.3~K and the spatial inhomogeneity in the overall $T_c$ increases to over 300~mK, although local (micron-scale) $T_c$ remains sharp. All scans were taken with a field coil ac current amplitude of $I_{\text{fc}} = $0.5~mA }
  \label{fig:Tseries_scans}
\end{figure*}

We combine scanning SQUID microscopy with a piezoelectric based strain apparatus similar to that described previously \cite{hicks2014piezoelectric}\cite{watson2018micron}. A schematic of the measurement setup is shown in Fig. \ref{fig:methods}(c). The sample is mounted with an open-face configuration where only one face of the sample (opposite of the scan surface) is anchored at each end to the movable titanium sample plates. In contrast, previous strain measurements employed a clamping configuration \cite{hicks2014piezoelectric}\cite{watson2018micron}. This open-face design allows for uninhibited access to the top face of the sample, which is beneficial for room temperature alignment between the sample and our SQUID susceptometer. This asymmetric mounting configuration between the top and bottom surfaces of the sample causes the sample to bend when the sample plates are displaced. To reduce this bending, our setup incorporates titanium flexures that resist sample bending and improve strain transfer to the top surface of the sample. We also note that because our SQUID susceptometer is a local probe, the sharpness of our observed transitions is determined by the local strain gradient within our measurement volume. Finite element simulations (see Appendix \ref{sec:appendix_strain_sims}) of the sample and strain device suggest that the magnitude of the strain gradient over the measurement volume are $\sim10^{-4}$\%. Thus, our susceptometers do not lose strain resolution because of long-length-scale strain gradients caused by bending. The sample is anchored to the titanium sample plates and flexures using Masterbond EP29LPSP epoxy and is oriented to apply uniaxial strain along the $\langle100\rangle$ crystal axis with a nominal strained sample length of 1.6 mm. Displacement of the sample plates is measured in situ using a parallel plate capacitor incorporated in the strain apparatus.


\section{Results and Discussion}
\label{sec:results}

\begin{figure}
 \centerline{\includegraphics[width=\linewidth]{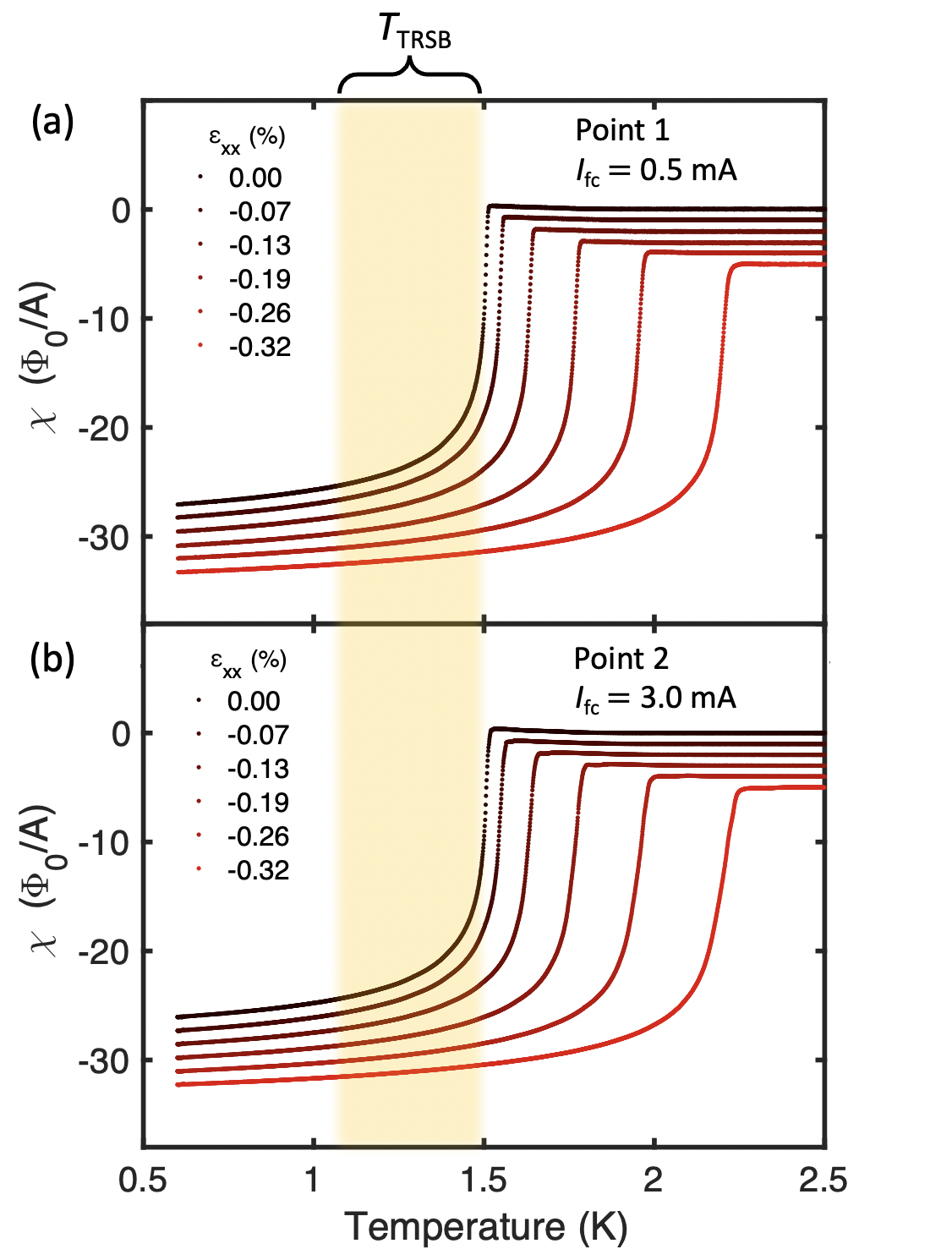}}
  \caption{Local susceptibility as a function of temperature recorded under different values of compressive strain. (a) and (b) Data taken at points 1 and 2 respectively as indicated in Fig. \ref{fig:Tseries_scans}(a). The yellow bands indicate the temperature range where an onset of TRSB has been observed in $\mu$SR measurements on uniaxially strained samples\cite{grinenko2021split}. All curves are offset vertically for clarity.}
  \label{fig:SUSC_T}
\end{figure}

Temperature series scans of susceptibility with the sample held under zero strain are shown in Fig. \ref{fig:Tseries_scans}(a)-(d).  At the lowest temperature $T = 1.0$ K, the sample shows almost uniform diamagnetism. Spots labeled 1 and 2 indicate where susceptibility versus temperature traces are recorded under a series of applied strains. At temperatures near the bulk $T_c=$1.5~K [Fig. \ref{fig:Tseries_scans}(b,c)], the diamagnetism becomes inhomogeneous due to spatial variation of local $T_c$. At the highest temperature [Fig. \ref{fig:Tseries_scans}(d)], small spots remain weakly diamagnetic, likely due to inclusions and defects inducing a local strain enhancement of $T_c$ \cite{maeno1998enhancement}\cite{kittaka2009spatial}. From these scans taken under zero applied strain, the spatial variation of the $T_c$ over this region of the sample is $\sim$50~mK. Figures \ref{fig:Tseries_scans}(e)-(h) show susceptibility scans over the same region while applying a compressive strain of $\epsilon_{xx}= -0.32\%$. The bulk $T_c$ of this region increases to approximately 2.3~K and the scale of the spatial inhomogeneity in $T_c$ increases to over 300~mK. Similar transition broadening under strain has been observed in previous bulk measurements \cite{hicks2014strong,steppke2017strong,li2021high} and is expected because the quadratic dependence of $T_c$ on strain implies that strain disorder increases variations in $T_c$ at values of $\epsilon_{xx}$ for which $dT_c/d\epsilon_{xx}$ is large.

Figures \ref{fig:SUSC_T} (a) and (b) show the susceptibility versus temperature traces under a series of fixed strains at point 1 ($I_{\text{fc}} = $0.5~mA) and point 2 ($I_{\text{fc}} = $3.0~mA) respectively. In this measurement, the height of the field coil from the sample is 0.4-0.6~$\mu$m and the field at the sample per unit of current through the field coil is 4-6 G/mA. The yellow shaded bands indicate the range of $T_{\text{TRSB}}$ that has been observed in $\mu$SR measurements under uniaxial strain \cite{grinenko2021split}. A notable feature of these temperature traces is the sharpness of the transitions at the highest applied strain which highlights the advantage of our local probe technique: while the large-scale spatial inhomogeneity in $T_c$ under strain is broadened to $\sim$300~mK [Fig. \ref{fig:Tseries_scans}(e)-(h)], the onset of diamagnetic screening over the local measurement volume ($\sim$10~$\mu$m$^3$) is rounded to less than $50$~mK [Fig. \ref{fig:SUSC_T}(a,b)]. Therefore, our experimental detection limit of a second transition is less constrained by transition broadening under compression as compared to bulk measurements. 

\begin{figure}
 \centerline{\includegraphics[width=\linewidth]{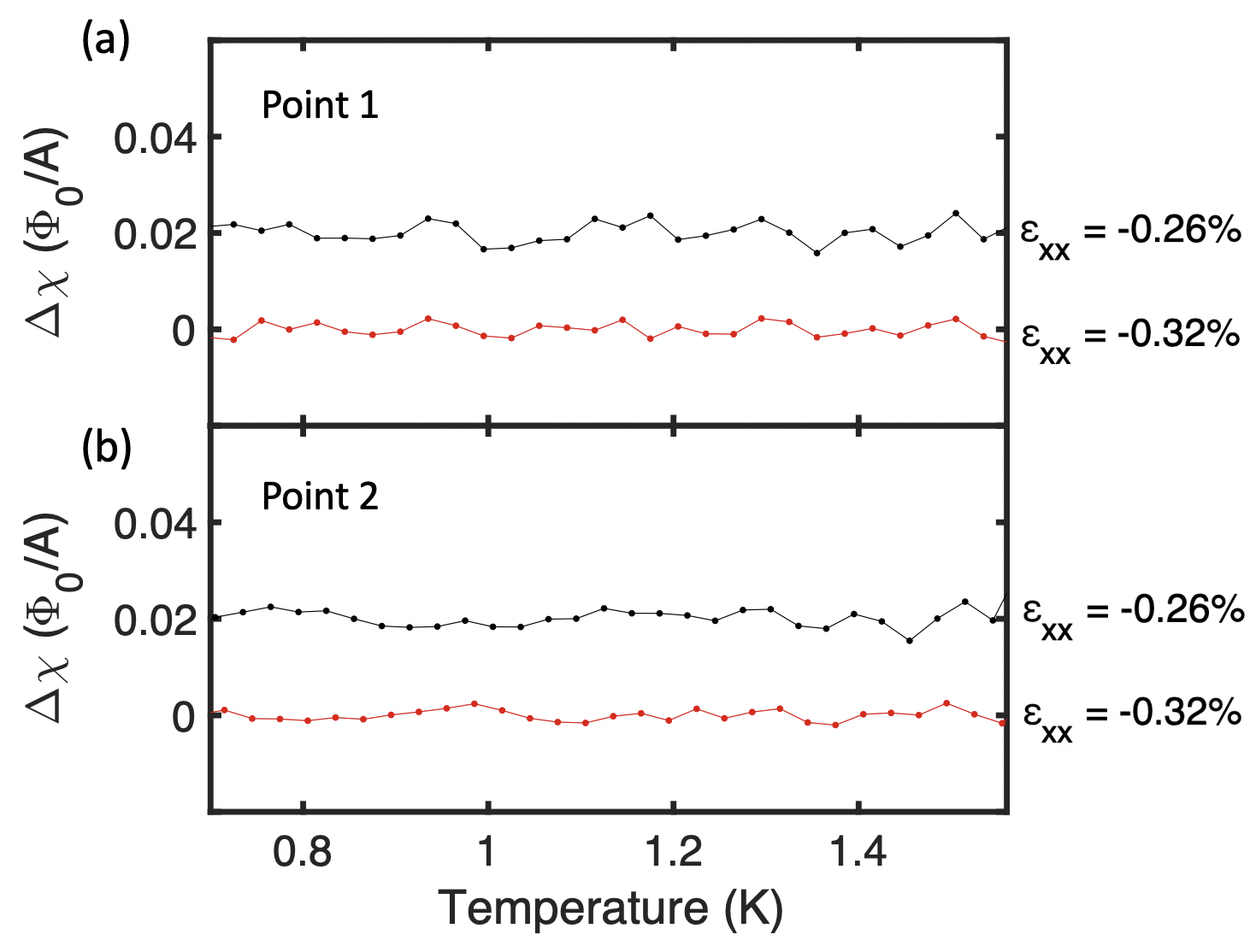}}
  \caption{Residuals of a $4^{\text{th}}$-order polynomial fit of the local susceptibility versus temperature data for applied strains of -0.26\% and -0.32\% over the temperature range 0.75-1.55K. (a) and (b) Results from data measured at points 1 and 2 respectively as indicated in Fig. \ref{fig:Tseries_scans}(a).  Curves are offset vertically for clarity}
  \label{fig:residuals}
\end{figure}

 None of the curves under compression show an obvious anomaly in the susceptibility below the initial onset of Meissner screening. To analyze the data with greater sensitivity, we average the data into bins of width $\Delta T= 30$~mK and subtract a fourth-order polynomial fit over the temperature range $T = $ 0.75-1.55~K. Figure \ref{fig:residuals} shows the residuals, $\Delta \chi$, after subtracting the polynomial fit for the datasets measured at the two highest applied strains of -0.26\% and -0.32\%. These curves show no evidence of a second transition. The rapidly increasing slope of $\chi(T)$ approaching $T_c$ [Fig.\ref{fig:SUSC_T}], which reflects the fact that $\lambda(T)$ diverges near $T_c$, makes it difficult to extend this analysis to $T>$1.55~K. Analysis of our experimental resolution (see Appendix \ref{sec:appendix_limit}) suggests that if there is a second transition over this temperature range, it contributes less than 1\% of the zero temperature superfluid density. 

 Our results, together with the non-observation of a second transition in heat capacity \cite{li2021high}, strongly suggest that there is no second transition under strain of the superconducting order parameter over the temperature range indicated by $\mu$SR measurements. The mechanism of the TRSB observed in Sr$_2$RuO$_4$ is still unclear. We note that previous scanning SQUID magnetometry measurements on Sr$_2$RuO$_4$ \cite{hicks2010limits}\cite{kirtley2007upper} as well as the candidate chiral $d$-wave superconductor URu$_2$Si$_2$ \cite{iguchi2021local} did not observe the spontaneous magnetization that would be expected for chiral superconductivity. Although Meissner screening and surface effects may suppress this spontaneous magnetization \cite{iguchi2021local}, the present results lend additional support to the hypothesis that the TRSB observed in Sr$_2$RuO$_4$ represents a separate transition that is not related to superconductivity.
 
 Another possibility is that the TRSB is related to superconductivity, but does not cause a large anomaly in the superfluid density. Intuitively, the change in penetration depth at $T_{TRSB}$ is weaker in cases where the two components share similar nodal structure over the Fermi surface. The even parity, orbital antisymmetric, spin-triplet model \cite{suh2020stabilizing} may be consistent with the observed absence of a clear change of the penetration depth. According to that model, the vertical gap minima remain near-nodes above $T_{c,2}$.


A non-observation of a second transition in the superfluid density, taken by itself, is most easily explained by a single-component order parameter. The elastocaloric measurements have recently been interpreted as ruling out any order parameter with vertical line nodes at the van Hove lines \cite{palle2023constraints}. Among the one-dimensional representations (see Table \ref{tab:table1}), this result allows only for $s$-wave or $d_{x^2-y^2}$-wave pairing. Evidence for vertical line nodes from scanning tunneling microscopy support the hypothesis of single-component $d_{x^2-y^2}$-wave pairing. However, a single-component order parameter would require reinterpretation of the experimental evidence for TRSB and both ultrasoud measurements that suggested a two-component order parameter. A possible explanation put forward in Ref. \cite{willa2021inhomogeneous} is a $d_{x^2-y^2}$-wave order parameter, in which an inhomogeneous second order parameter is induced locally around sample dislocations.

\section{Conclusion}

In summary, we measure the temperature dependence of the magnetic susceptibility of Sr$_2$RuO$_4$ under uniaxial strain using scanning SQUID microscopy. By using a local probe with a measurement volume of $\sim$10~$\mu$m$^3$, we resolve sharp superconducting transitions that remain sharp ($\Delta T_c<50$~mK) up to our highest applied strain. As a result, our experimental detection limit of a second transition is less constrained by transition broadening due to strain inhomogeneity than bulk measurements. Our results show no evidence of a strain-induced splitting of a two-component order parameter. We place an upper limit of less than 1\% on the contribution to the zero temperature superfluid density from a second order parameter with a transition in the temperature range suggested by $\mu$SR data, suggesting that such a transition does not occur. This result provides further constraints on theories of a two-component order parameter in Sr$_2$RuO$_4$.

\begin{acknowledgments}

This work is supported by the Department of Energy, Office of Basic Energy Sciences, Division of Materials Sciences and Engineering, under contract DE-AC02-76SF00515. Y.M. was supported by JSPS KAKENHI (Grant No. JP22H01168)

\end{acknowledgments}

\appendix

\section{Experimental limit on the detection of a second transition in susceptibility}
\label{sec:appendix_limit}

Our SQUID susceptometers are sensitive to changes in the London penetration depth, $\lambda$. Therefore, our experiment will not resolve a second transition, $T_{c,2}$, if the signature in $\lambda$ associated with $T_{c,2}$ is smaller than our experimental resolution. Here, we estimate our experimental resolution and thereby place an upper limit on an anomaly in the superfluid density.

We start with a phenomenological two-fluid model in which the superfluid density is given by \footnote{The two-fluid model can alternatively be expressed with a prefactor of $(1-\Delta n_s)$ on the $T_{c,1}$ term such that the superfluid density is independent of $\Delta n_s$ at $T = 0$~K. For clarity, we chose a modified form for the two-fluid model to explicitly indicate the extra zero temperature superfluid density added by the second transition. The choice of the two-fluid model does not significantly impact the results of the analysis.},

\begin{equation}
\frac{\lambda^2(0)}{\lambda^2(T)} =  \left(1-\frac{T^4}{T_{c,1}^4}\right) + \Delta n_s\left(1-\frac{T^4}{T_{c,2}^4}\right)
\label{eq:lambda_T}
\end{equation}
 where $T_{c,1}$ and $T_{c,2}$ correspond to the upper and lower superconducting transitions respectively and $\Delta n_s$ corresponds to the fraction of the $T = 0$~K superfluid density associated with the second transition at $T_{c,2}$. Taking $\Delta n_s=$ 0\% reduces to the case of a single superconducting transition at $T_{c,1}$. 

\begin{figure}
 \centerline{\includegraphics[width=\linewidth]{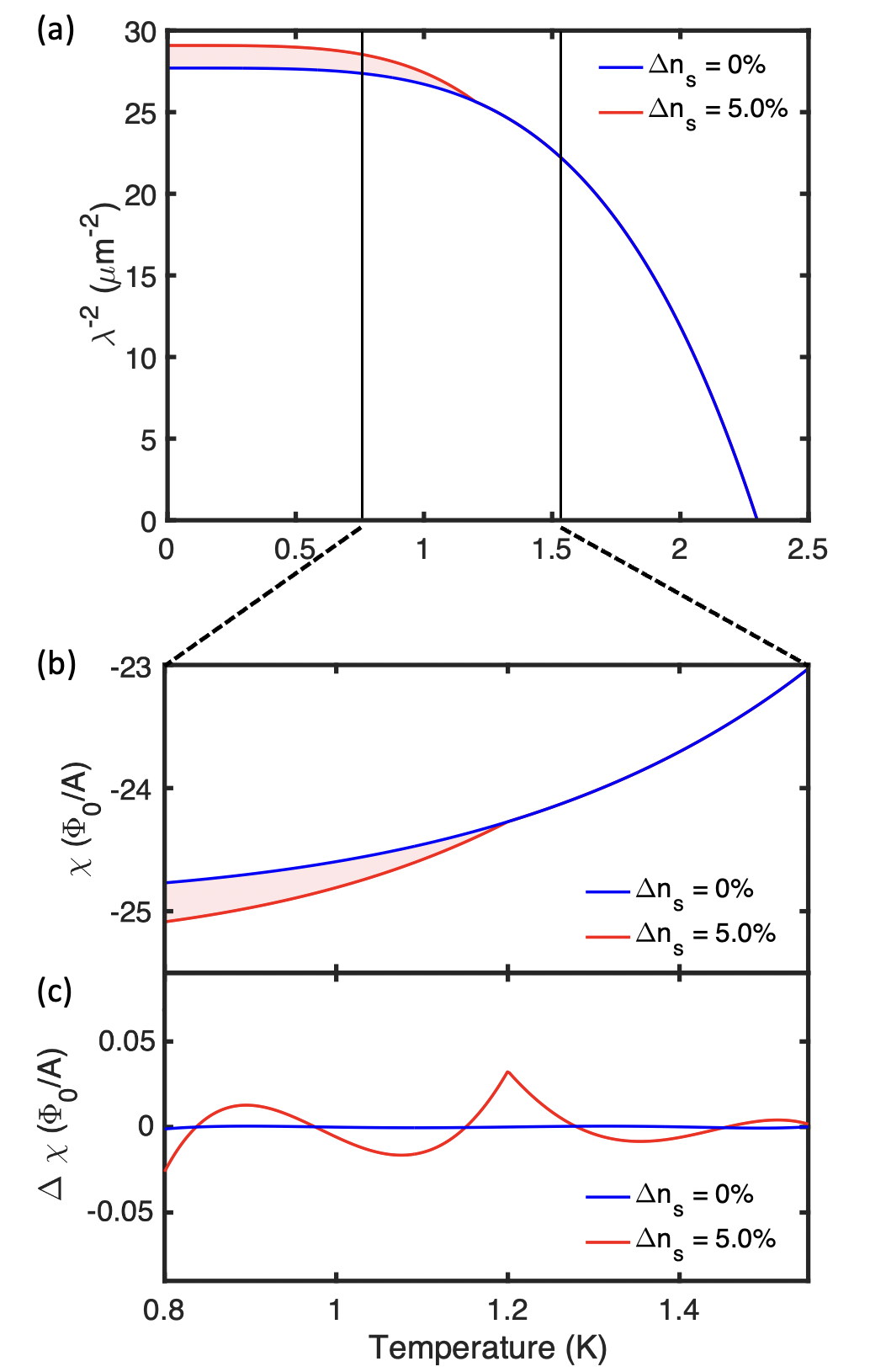}}
  \caption{Model of second transition in susceptibility (a) Superfluid density versus temperature for a single superconducting transition (blue curve) and a double transition (red dashed curve) (b) Modeled susceptbility for a single transition (blue curve) and for a double transition over the temperature range near $T_{c,2}$ (c) Residuals of 4$^{th}$-order polynomial fit for single transition (blue curve) and double transition (red curve).}
  \label{fig:superfluid_kink}
\end{figure}
 
 Figure \ref{fig:superfluid_kink}(a) shows the temperature dependence of the superfluid density, $\lambda^{-2}(T)$,  for the model scenario of a single superconducting transition (blue curve) at $T_{c,1} = $ 2.3~K. The red curve shows the scenario with a second transition at $T_{c,2} = $ 1.2~K which adds $\Delta n_s=$ 5\% to the $T = 0$ K superfluid density. In this model we set $\lambda(0) = $190~nm \cite{bonalde2000temperature}. The second transition at $T_{c,2}$ produces a kink in the superfluid density as a function temperature which is highlighted by the pink shaded area of Fig. \ref{fig:superfluid_kink}(a) .

\begin{figure}
 \centerline{\includegraphics[width=\linewidth]{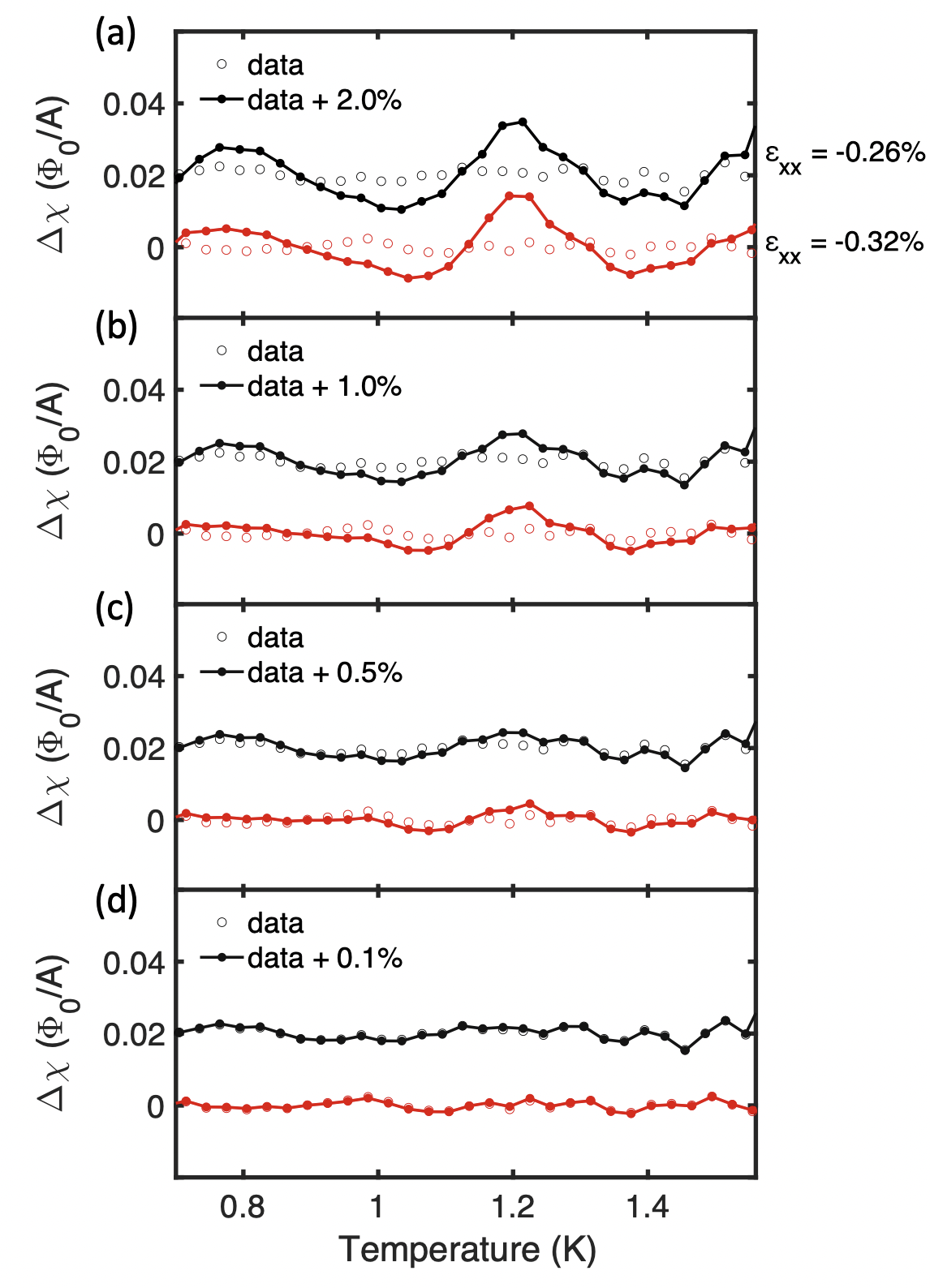}}
  \caption{Modeled signature of a second transition in susceptibility for varying amplitudes of $\Delta n_s$. (a)-(d) Plot of residuals, $\Delta\chi$, versus temperature after a model of a second transition was added to the raw data measured at point 2 with $T_{c,2} = $1.2~K and magnitude (a) $\Delta n_s = $ 2.0\%, (b) $\Delta n_s = $ 1.0\%, (c) $\Delta n_s = $ 0.5\%, and (d) $\Delta n_s = $ 0.1\%. Residuals with no second transition added (open circles) are shown for comparison. Results from dataset taken at applied strain of -0.26\% (-0.32\%) are shown in black (red). Curves at different strain values are offset vertically for clarity.        }
  \label{fig:susc_kink_limit}
\end{figure}

We next approximate the response of our SQUID susceptibility measurement resulting from this kink in the superfluid density at $T_{c,2}$. The dependence of $\chi$ on the local London penetration depth has been modeled for our SQUID susceptometers which approximates the pickup loop-field coil pair as thin circular wires \cite{kirtley2012scanning}. In the limit in which $\lambda \ll r$, $z$, where $r$ is the radius of the field coil and $z$ is the separation between the sample and field coil, the relation between $\lambda$ and $\chi$ can be modeled by

\begin{equation}
\lambda = \frac{r}{2}\left(\frac{1}{\bar{\chi}^{2/3}}-1 \right)^{1/2} - z.
\end{equation}

Here, $\bar{\chi} = \chi/\chi_0$, where $\chi_0$ is the mutual inductance of the field coil-pickup loop pair in the absence of a magnetically responsive sample and is experimentally determined to be $\chi_0 \approx 170$~$ \Phi_0/A$.
In this model, we use the temperature dependence of $\lambda$ given by Eq. (\ref{eq:lambda_T}) and we set the following parameters: $r =$ 800~nm and $z = $450~nm. Figure \ref{fig:superfluid_kink}(b) shows the modeled susceptibility versus temperature curves over the temperature range $T = $~0.8-1.55~K for the single transition (blue curve) and double transition (red curve) scenarios from Fig. \ref{fig:superfluid_kink}(a). The pink shaded area represents the 'extra' signal in $\chi$ resulting from the increase in superfluid density at the second transition $T_{c,2} = $ 1.2~K.  We note this model is approximate as the field coil has a slightly non-circular shape. Nevertheless, our experimentally observed magnitude of $\chi$ over this temperature range is well approximated by this model for a reasonable choice of the parameters $\lambda(0)$, $r$, and $z$. Figure \ref{fig:superfluid_kink}(c) shows the residuals, $\Delta \chi$, after subtracting a fourth-order polynomial fit to the curves in Fig. \ref{fig:superfluid_kink}(b). For the case of a single transition (blue curve), $\Delta \chi$ remains nearly zero over the temperature range shown while a double transition (red curve) causes a sharp cusp at $T_{c,2}$.

We add the 'extra' susceptibility signal modeled for a second transition (pink shaded region in Fig. \ref{fig:superfluid_kink} (b)) to our raw data and determine at which value of $\Delta n_s$ the cusp at $T_{c,2}$ is no longer visible in a plot of the residuals, $\Delta \chi$, versus temperature [Fig. \ref{fig:susc_kink_limit}]. For comparison, the open circles in Fig. \ref{fig:susc_kink_limit} show $\Delta \chi$ with no second transition added and are identical to the data shown in Fig. \ref{fig:residuals}(b). For $\Delta n_s =$ 2.0\% [Fig. \ref{fig:susc_kink_limit}(a)], a cusp at $T_{c,2}=$ 1.2~K is clearly visible. In our judgement, a cusp is still visible with $\Delta n_s =$ 1.0\% [Fig. \ref{fig:susc_kink_limit} (b)], but is no longer visible for $\Delta n_s =$ 0.5\% or lower [Fig. \ref{fig:susc_kink_limit} (c) and (d)]. Therefore, our results rule out any second transition below $T_c$ that contributes more than 1\% of the zero temperature superfluid density.

\section{Finite element strain simulation}
\label{sec:appendix_strain_sims}

 Achieving high strain homogeneity is a significant challenge in uniaxial pressure experiments. The asymmetric mounting geometry in which one face of the sample is bonded at each end to titanium sample plates offers uninhibited access to the top face of the sample, but has been shown to cause sample bending and result in high strain inhomogeneity under pressure \cite{hicks2014piezoelectric}. To reduce this bending, our sample plates include titanium flexures that resist bending without attenuation of the applied stress. Here, we present results of finite element strain simulations and compare the strain inhomogeneity induced in the sample when bonded to sample plates with and without the titanium flexures.

  \begin{figure}
 \centerline{\includegraphics[width=\linewidth]{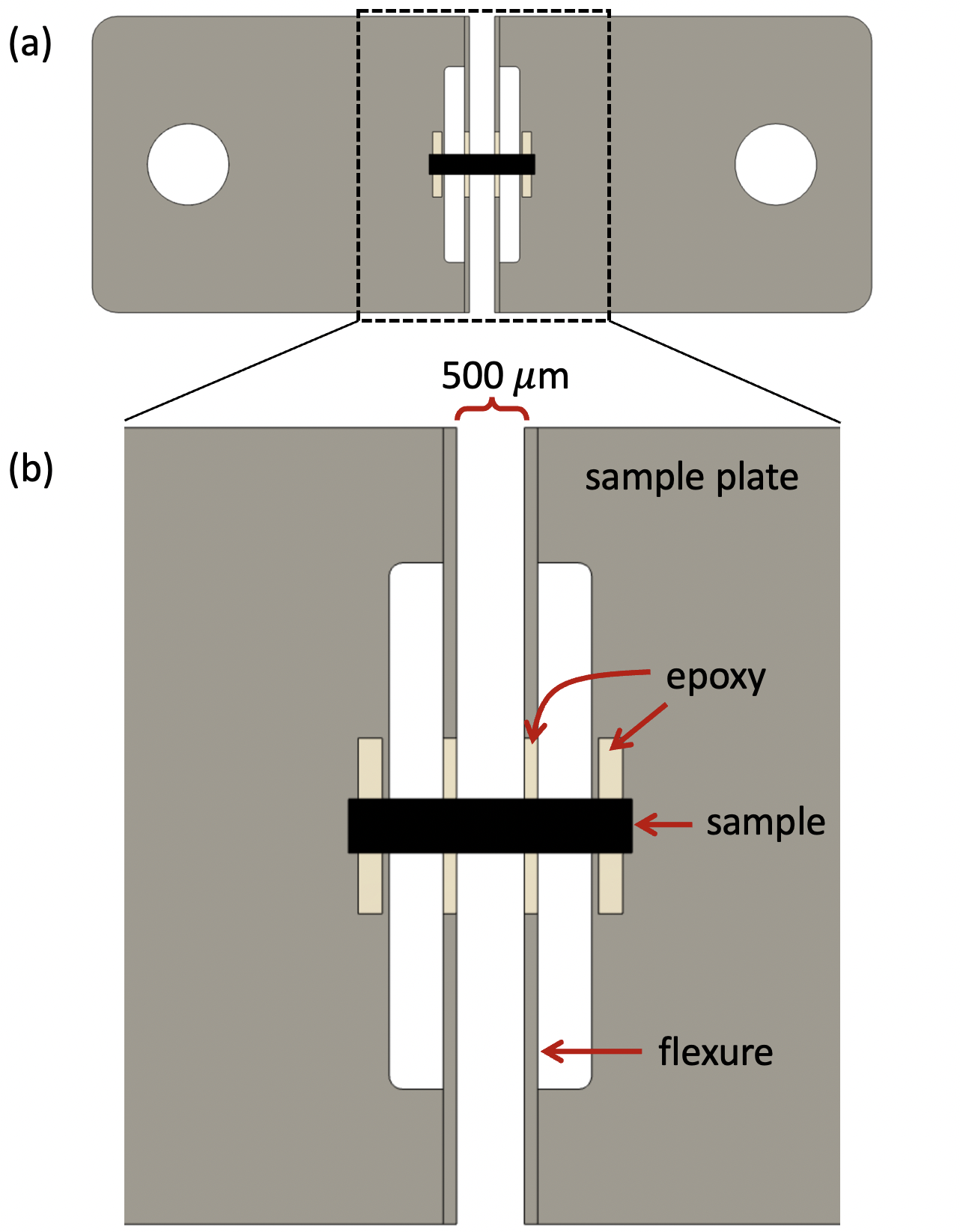}}
  \caption{Sample mounting setup (a) Top view of the model used in the finite element stress analysis  (b) Zoomed in view around the sample and titaniun flexures}
  \label{fig:strain-setup}
\end{figure}

Figure \ref{fig:strain-setup}(a) shows a schematic of the model used in the analysis and Fig. \ref{fig:strain-setup}(b) shows a zoomed in view of the setup around the sample and flexures. The bottom face of the sample is bonded to the titanium sample plates and flexures with a thin layer of epoxy. For simplicity, we assume both the sample and the epoxy to be isotropic with a a Poisson's ratio of 0.3. For Sr$_2$RuO$_4$, we used a Young's modulus of 176 GPa \cite{paglione2002elastic}. The elastic properties of Masterbond EP29LPSP have not been measured in the temperature range of our experiment. We therefore choose to set the Young's modulus of the epoxy to be 15 GPa which is consistent with previous studies using the low elastic epoxy Stycast 2850~\cite{hicks2014piezoelectric}\cite{hicks2014strong}. The sample thickness was taken to be $t=$70~$\mu$m. The strained length of the sample was $L = $1.6~mm. The stress is transferred to the sample via shear stresses in the epoxy layer. Typically, an epoxy thickness of a few tens of $\mu$m is desired to avoid failure of the epoxy due to buildup of shear stresses in the epoxy layer \cite{hicks2014piezoelectric}. While preparing the sample on the strain cell for the experiment, 19 $\mu$m diameter nylon wires were placed between the sample and the titanium sample plate to ensure sufficient epoxy thickness. Therefore, we set the epoxy thickness to be 20$\mu$m in our model.

\begin{figure}
 \centerline{\includegraphics[width=\linewidth]{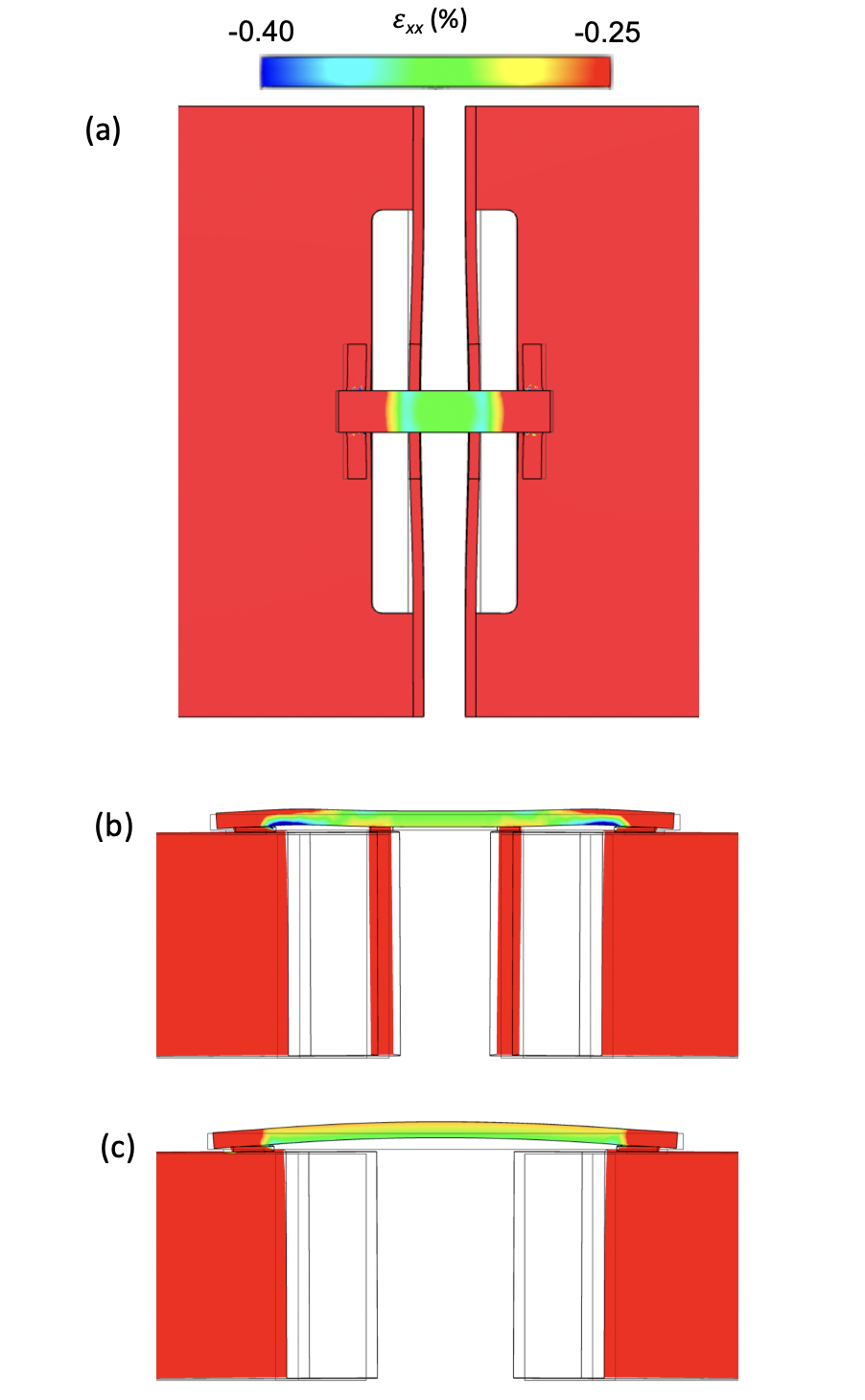}}
  \caption{Finite element strain simulations of the sample under compression. (a) Simulated $\epsilon_{xx}$ for the model with flexures viewed from above the sample (b) Side view cross section of the simluation shown in (a) cut along the length of the sample (c) Simulated $\epsilon_{xx}$ for the model without flexures. Deformation in all figures  is exaggerated by a factor of 10. Finite element simulations were performed using Autodesk Fusion 360 }
  \label{fig:strain-sims}
\end{figure}

Figure \ref{fig:strain-sims}(a) shows the simulated $\epsilon_{xx}$ strain in the system for a displacement of the movable sample plates from the zero strain position of 0.6\% of $L$. Deformation of the titanium flexures occurs as the sample is compressed. Figure \ref{fig:strain-sims}(b) shows the $\epsilon_{xx}$ strain for a cross section along the length of the sample. We note that as the sample plates are displaced, the load applied to the sample through the epoxy is transferred over a non-zero length scale which depends on parameters such as the thickness or Young's modulus of the epoxy. This load transfer length increases the effective strained length of the sample and as a result, the actual strain in the sample is lower than 0.6\%. Slight buckling of the sample can be seen in the region near where the sample is epoxied to the sample plates. For comparison, Fig. \ref{fig:strain-sims}(c) shows the simulation results for the scenario with the titanium flexures removed. For the same displacement of the sample plates, the setup without the flexures increases the sample bending. The case with flexures generates a $\sim$20\% increase in the maximum applied strain at the top surface because bending decreases the strain transferred to the top surface of the sample.

From these strain simulations, we can also estimate the magnitude of the strain gradient over the volume probed by our our SQUID susceptometer. The strain gradient between the top and bottom faces of the sample is $\Delta\epsilon_{xx}=0.02$\%  and the gradients in the in-plane direction are negligible (Fig. \ref{fig:strain-sims} (b)). Our measurement is sensitive to length scales of order the diameter of the field coil, $D\sim1$~$\mu$m, or $\sim$1\% of the sample thickness. We estimate a strain gradient over our measurement volume of $(D/t)\Delta\epsilon_{xx} \approx 2\times10^{-4}$\%, which highlights the advantage of our local probe technique.

\bibliography{SRO}

\end{document}